\documentclass[aps,prl,twocolumn,superscriptaddress,notitlepage,nofootinbib,showkeys,showpacs]{revtex4-1}
\usepackage{amsmath}\allowdisplaybreaks[1]
\usepackage{physics, bm}
\usepackage{graphicx}
\usepackage{amssymb}
\usepackage[colorlinks, linkcolor=red, citecolor=red, urlcolor=magenta]{hyperref}

\usepackage{xcolor}    
\usepackage[normalem]{ulem}
\usepackage{multirow}
\usepackage{etoolbox}
\usepackage{orcidlink}

\newcommand{\beq}{\begin{equation}}
\newcommand{\eeq}{\end{equation}}
\newcommand{\bea}{\begin{eqnarray}}
\newcommand{\eea}{\end{eqnarray}}
\newcommand{\nn}{\nonumber}

\newcommand{\M}{\mathcal{M}}

\makeatletter
\patchcmd\frontmatter@PACS@format{\addvspace{11\p@}}{}{}{}
\pretocmd\frontmatter@keys@format{\addvspace{11\p@}}{}{}
\patchcmd{\titleblock@produce}
  {\@pacs@produce\@pacs\@keywords@produce\@keywords}
  {\@keywords@produce\@keywords\@pacs@produce\@pacs}
  {}{}
\makeatother

\usepackage{lipsum}

\begin{document}

\preprint{
	{\vbox {		
		\hbox{\bf JLAB-THY-24-4136}	
		\hbox{\bf MSUHEP-24-009}
}}}
\vspace*{0.2cm}

\title{Dihadron azimuthal asymmetry and light-quark dipole moments \\at the Electron-Ion Collider}

\author{Xin-Kai Wen\,\orcidlink{0009-0008-2443-5320}}
\affiliation{School of Physics, Peking University, Beijing 100871, China}
\affiliation{Institute of High Energy Physics, Chinese Academy of Sciences, Beijing 100049, China}
\affiliation{China Center of Advanced Science and Technology, Beijing 100190, China}

\author{Bin Yan\,\orcidlink{0000-0001-7515-6649}}
\email{yanbin@ihep.ac.cn}
\affiliation{Institute of High Energy Physics, Chinese Academy of Sciences, Beijing 100049, China}
\affiliation{Center for High Energy Physics, Peking University, Beijing 100871, China}

\author{Zhite Yu\,\orcidlink{0000-0003-1503-5364}}
\email{zyu1@bnl.gov}
\affiliation{Theory Center, Jefferson Lab, Newport News, Virginia 23606, USA}
\affiliation{High EnergyTheory Group, Physics Department,Brookhaven National Laboratory, Upton, NY11973, USA}

\author{C.-P. Yuan\,\orcidlink{0000-0003-3988-5048}\,}
\affiliation{Department of Physics and Astronomy, Michigan State University, East Lansing, Michigan 48824, USA}


\begin{abstract}
We propose a novel method to probe light-quark dipole moments by examining the azimuthal asymmetries between a collinear pair of hadrons in semi-inclusive deep inelastic lepton scattering off an unpolarized proton target at the Electron-Ion Collider. These asymmetries provide a means to observe transversely polarized quarks, which arise exclusively from the interference between the dipole and the Standard Model interactions, thereby depending linearly on the dipole couplings. We demonstrate that this novel approach can enhance current constraints on light-quark dipole operators by an order of magnitude, free from contamination of other new physics effects. Furthermore, it allows for a simultaneous determination of both the real and imaginary parts of the dipole couplings, offering a new avenue for investigating potential $CP$-violating effects at high energies.
\end{abstract}


\maketitle

\section{ Introduction}

Dipole moments are intrinsic quantum properties of particles,
characterizing chirality-flip effects in their interactions with gauge fields.
Precise calculations and measurements of these fundamental quantities serve as stringent tests for the Standard Model (SM) of particle physics
and provide insights into potential new physics (NP) effects beyond the SM. 
A notable example involves the anomalous magnetic dipole moments (MDMs) and electric dipole moments (EDMs) of leptons~\cite{Belle:2002nla,DELPHI:2003nah,Hanneke:2008tm,Muong-2:2023cdq,Shao:2023bga,Ma:2024fmy}
related to their interactions with photons.
In particular, the recent muon MDM measurement at Fermilab has propelled a number of studies 
aiming to reveal potential NP residing in the deviation of experimental data from the SM prediction~\cite{Muong-2:2023cdq}. 
At high energies, EDMs and MDMs combine with weak dipole moments associated with the $W$ and $Z$ bosons
into a unified electroweak structure,
as parametrized by dimension-six operators in the Standard Model Effective Field Theory (SMEFT)~\cite{Buchmuller:1985jz,Grzadkowski:2010es}.
These electroweak (EW) dipole interactions provide new helicity structures in high-energy scattering
and could well be the source for some anomalies in the current experimental data,
such as the long-standing anomaly of Lam-Tung relation breaking observed in the Drell-Yan process at the Large Hadron Collider (LHC)
as suggested recently~\cite{Li:2024iyj,Li:2025fom}.

The EW dipole couplings are notoriously difficult to be constrained in a global fit of the SMEFT parameters~\cite{Beneke:2014sba,Craig:2015wwr,Ge:2016zro,Khanpour:2017cfq,Chiu:2017yrx,Durieux:2017rsg,Barklow:2017suo,Li:2019evl,Rao:2019hsp,Yan:2021tmw}.
The reason is that dipole interactions flip fermion helicities, so their interference with the SM amplitudes 
at $\order{1/\Lambda^2}$ is zero for light fermions due to the chiral symmetry, if one only considers unpolarized cross sections,
in which the dipole contributions are suppressed by $\order{1/\Lambda^4}$, with $\Lambda$ representing the NP scale. 

On the other hand, if the fermion under study has a single transverse spin, 
a nonzero interference between the dipole and SM interactions can be generated at $\order{1/\Lambda^2}$~\cite{Wen:2023xxc, Boughezal:2023ooo, Wang:2024zns}.
As demonstrated in Ref.~\cite{Wen:2023xxc}, at a future high-energy electron-positron collider,
a single transversely polarized beam can induce an azimuthal asymmetry in the overall event distributions.
A precision measurement of this asymmetry can significantly improve the constraints on the electron EW dipole moments.
A similar conclusion can be extended to a future muon collider, if the muon beams can be transversely polarized~\cite{AlAli:2021let, Accettura:2023ked}.

However, this strategy is not applicable  to light quarks  because of  quark confinement in Quantum Chromodynamics,
which prevents a direct measurement of the quark EW dipole moments.
One potential avenue is through the nucleon dipole moment measurements,
but interpreting the results in terms of quark constituents inevitably relies on the nonperturbative input of the nucleon spin structure~\cite{Bicudo:1998qb,Hecht:2001ry,Bhattacharya:2012bf,Pitschmann:2014jxa,Graner:2016ses,Sahoo:2016zvr,Schneider:2017lff,Liu:2017olr,Abel:2020pzs,Tiesinga:2021myr,Kawaguchi:2022dbq,ParticleDataGroup:2022pth,Mena:2024qou,Wu:2024xpq,Han2026ImprovingMC}.
Within collinear factorization, transversely polarized quarks in the initial state can only come from transversely polarized nucleon beams 
through the nonperturbative transversity distribution $h_q(x, \mu)$, which can constrain the quark dipole moments at the upcoming Electron-Ion Collider (EIC)
in a way similar to constraining the dipole moments of leptons~\cite{Boughezal:2023ooo, Wang:2024zns}.
The effects depend on both the degree of transverse polarization of the incoming beams and the quark transversity distributions.

In this work, we focus on unpolarized nucleon scattering and probe the quark dipole moments through 
the transverse spin of a {\it final}-state quark.
Detectors at collider experiments do not directly measure spins. However, for an unstable particle that decays, spin information is encoded in the angular distribution of its decay products,
which can be analyzed to deduce the mother particle's spin~\cite{Kane:1991bg,Yu:2021zmw}.
Now, a light quark does not decay, but fragments into a jet of hadrons due to color interactions after being produced from a hard scattering.
The transverse spin $\bm{s}_{T, q}$ then dictates an azimuthal asymmetry of the jet constituents with respect to the quark momentum direction, for it breaks the rotational symmetry.
To reveal this asymmetry at the hadron level requires to measure at least two hadrons $h_1(p_1)$ and $h_2(p_2)$ in the jet, 
which form a plane with a preferential orientation with respect to $\bm{s}_{T, q}$,
controlled formally by the unique scalar $\bm{s}_{T, q} \cdot (\bm{p}_1 \times \bm{p}_2)$. 

\section{Theoretical formalism} 
Based on above analysis, we propose to measure the final-state light quark transverse spin effects resulting from the EW dipole interactions in the two-hadron production in semi-inclusive deep inelastic scattering (SIDIS),
\beq\label{eq:sidis}
	e^-(\ell) + p(p) \to e^-(\ell^\prime) + h_1(p_1) + h_2(p_2) + X,
\eeq
where the electron exchanges with the proton a highly virtual photon $\gamma^*$ or $Z$ boson of momentum $q = \ell - \ell'$.
As usual, we work in the Breit frame, where the $\gamma^*$ or $Z$ moves along the $\hat{z}$ direction 
and hits the proton at rest to produce two hadrons $h_1(p_1)$ and $h_2(p_2)$ plus anything else (collectively denoted by $X$),
and define the kinematic variables,
\begin{equation*}
	Q^2 = -q^2, \quad 
	x = \frac{Q^2}{2p\cdot q}, \quad 
	y = \frac{p\cdot q}{p\cdot \ell}, \quad 
	z = \frac{p\cdot P_h}{p\cdot q},
\end{equation*}
with $P_h = p_1 + p_2$ and $M_h^2 = P_h^2$ being the total momentum and invariant mass of the two hadrons, respectively.

We focus on the kinematic region where $M_h \ll |\bm{P}_h|$, such that
the leading contribution to the cross section of Eq.~\eqref{eq:sidis} comes from region where the two hadrons come from a single parton fragmentation.
This results in a factorization formula like the one-hadron production in SIDIS, but with the single-hadron fragmentation function replaced by 
two dihadron fragmentation functions (DiFFs), $D_{h_1 h_2 / q}$ and $H_{h_1 h_2 / q}$~\cite{Collins:1993kq, Artru:1995zu, Jaffe:1997hf, Jaffe:1997pv, Bianconi:1999cd, Bianconi:1999uc, 
Barone:2001sp, Bacchetta:2003vn, Bacchetta:2008wb, Zhou:2011ba, Metz:2016swz}, as
\begin{align}
		&\frac{d\sigma}{dx \, dy \, dz \, dM_h \, d\phi_R} = \frac{N}{2\pi} \sum_q  f_q(x, Q) \big[ D_{h_1 h_2 / q}(z, M_h; Q) 	\nn\\
		& \hspace{0.4em} 
			- ( \bm{s}_{T, q}(x, Q) \times \hat{\bm{R}}_T )^z H_{h_1 h_2 / q}(z, M_h; Q) \big]  
			C_q(x, Q),
\label{eq:diXsec}
\end{align}
where $N = y / (16\pi Q^2)$ and we have used the leading-order (LO) approximation to include only quarks and antiquarks
and set the initial-state quark momentum fraction to $x$ in the parton distribution function $f_q$ 
as well as the final-state dihadron momentum fraction to $z$ in the DiFFs $D$ and $H$.
At a moderate $z$, the dihadron momentum is of the same order as $Q$, which is the hard scale in this process and thus we set the factorization scale $\mu = Q$. 

\begin{figure}
	\centering
	\includegraphics[scale=0.4]{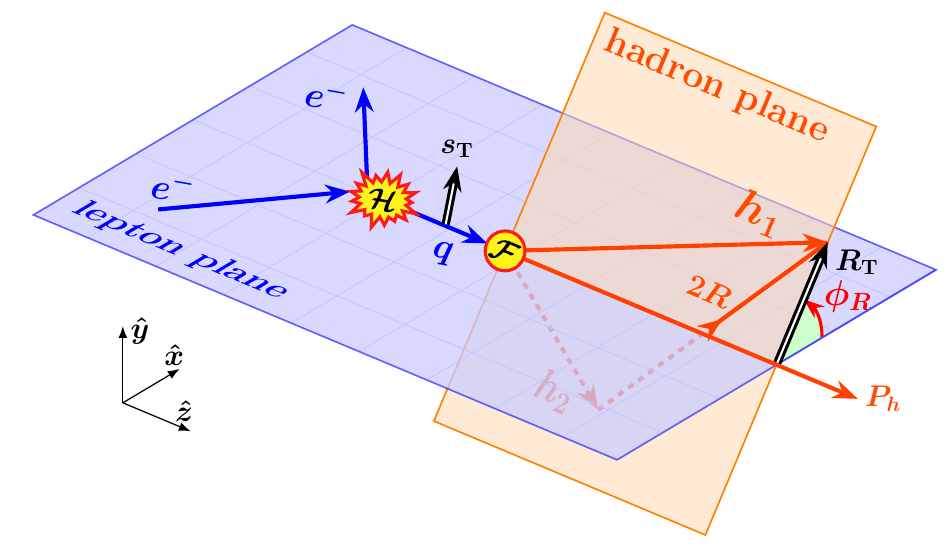}
	\caption{LO kinematic configuration of the dihadron production in SIDIS.
	The proton is at rest in $\mathcal{H}$ and is not displayed explicitly.
	$q$ is the struck quark that fragments into $h_1$ and $h_2$ (aside from anything else).
	}
	\label{Fig:Geo}
\end{figure}

The LO kinematics is depicted in Fig.~\ref{Fig:Geo},
\footnote{Beyond LO, one should first construct the hadron plane using $P_h$ and the electron plane (e.g., following the Trento convention~\cite{Bacchetta:2004jz}), 
and then build the dihadron plane which intercepts with the hadron plane by an angle $\phi_R$.}
where we also define the two transverse directions,
with the $\hat{x}$ axis lying on the electron scattering plane and the $\hat{y}$ axis perpendicular to it.
The final-state quark moves along $\hat{z}$ and fragments into a jet in which the two hadrons $(h_1 h_2)$ to be detected form a secondary hadron fragmentation plane.
These two planes intersect at the $\hat{z}$ axis and make an angle $\phi_R$, which is the azimuthal angle of the hadron pair's relative momentum $R = (p_1 - p_2) / 2$.

The factorization formula, cf. Eq.~\eqref{eq:diXsec}, separates the production and the fragmentation of the quark.
First, the production part is described by the hard electron-quark scattering 
$e_{\lambda_e} q_{\lambda_q} \to e_{\lambda'_e} q_{\lambda'_q}$, with $\lambda$'s referring to helicities.
The square of its amplitude $\M_{\lambda_q\lambda'_q}^{\lambda_e\lambda'_e}$
determines both the unpolarized production rate $C_q$ and the spin vector $\bm{s}_{q} = (s^x_{q}, s^y_{q}, s^z_{q})$ of the final-state quark through
\beq\label{eq:hard}
	W^q_{\lambda\lambda'} \equiv \frac{1}{4 N_c} \M_{\lambda_q \lambda}^{\lambda_e \lambda'_e} \big( \M_{\lambda_q \lambda'}^{\lambda_e\lambda'_e} \big)^*
	= \frac{C_q }{2} \big( \delta_{\lambda\lambda'} + s_q^i \, \sigma^i_{\lambda\lambda'}  \big),
\eeq
which plays the role of the (unnormalized) quark spin density matrix, and
where the repeated indices are summed over, initial-state spins and colors are averaged ($N_c = 3$ is the quark color number),
and $\sigma^i = (\sigma^x, \sigma^y, \sigma^z)$ are the Pauli matrices.
The final-state quark carries the spin $\bm{s}_{q}$ and fragments into hadrons, as described by two DiFFs.
The unpolarized DiFF $D_{h_1 h_2 / q}(z, M_h; Q)$ describes the probability of fragmenting into a hadron pair $(h_1 h_2)$ at the values of $z$ and $M_h$,
while the associated interference DiFF $H_{h_1 h_2 / q}(z, M_h; Q)$ 
modulates how the quark transverse spin $\bm{s}_{T, q}$ is transmitted into the hadronic asymmetry, with
\beq
	( \bm{s}_{T, q} \times \hat{\bm{R}}_T )^z = s^x_{q} \sin\phi_R - s^y_{q} \cos\phi_R.
\eeq
In other words, the ratio $H_{h_1 h_2 / q} / D_{h_1 h_2 / q}$ is the (differential) spin analyzing power.
We note that this simple interpretation relies on a particular convention adopted for the DiFF definition, as in Ref.~\cite{Cocuzza:2023vqs}.

Importantly, due to chiral symmetry, the transverse quark spin cannot be measured through a perturbative observable, unlike linear gluon polarization~\cite{Yu:2022kcj}. Therefore, we must introduce a nonperturbative probe, such as the dihadron azimuthal asymmetry (without measuring the hadron spins), which is in turn factorized into the nonperturbative but universal DiFFs that can be obtained by fitting to experimental data.
The nonzero interference DiFF $H_{h_1 h_2 / q}$ originates from the breaking of chiral symmetry,
and can have sizable values according to the global fit results~\cite{Bacchetta:2011ip,Radici:2018iag,Pitonyak:2023gjx,Cocuzza:2023oam,Cocuzza:2023vqs}.

The transverse spin $\bm{s}_{T, q}$ in Eq.~\eqref{eq:hard} is associated with the off-diagonal elements of the density matrix,
$(s_q^x, \, s_q^y) \propto (\Re W^q_{-+}, \, \Im W^q_{-+})$,
which require a single helicity flip for the quark in either the amplitude or the conjugate amplitude.
In the SM, this is forbidden by the gauge interactions or severely suppressed by the light quark mass~\cite{Wen:2023xxc, Boughezal:2023ooo, Wang:2024zns}.
Because of this, the value of $\bm{s}_{T, q}$ is highly sensitive to the quark dipole interactions in the SMEFT~\cite{Buchmuller:1985jz,Grzadkowski:2010es}, 
which we take as
\begin{align}
	&\mathcal{L}_{\rm eff} = -\frac{1}{\sqrt{2}} \bar{q}_L \sigma^{\mu\nu}
	\left(g_1 \Gamma_B^d B_{\mu\nu}+g_2 \Gamma_W^d \sigma^a W_{\mu\nu}^a \right) \frac{H}{v^2} d_R
	\nn\\
	&\hspace{1em} 
	-\frac{1}{\sqrt{2}} \bar{q}_L \sigma^{\mu\nu}
	\left(g_1 \Gamma_B^u B_{\mu\nu}+g_2 \Gamma_W^u \sigma^a W_{\mu\nu}^a \right) \frac{\widetilde{H}}{v^2} u_R
	+{\rm h.c.},
\label{eq:Leff}
\end{align}
where $q_L = (u_L, d_L)^T$ is the first-generation left-handed quark doublet and $u_R$ and $d_R$ are the right-handed singlets.
$H$ and $\widetilde{H} = i \sigma^2 H^*$ are the Higgs doublet fields, with $v=246~{\rm GeV}$ being the vacuum expectation value. 
The field strength tensors of the gauge groups $U(1)_Y$ and $SU(2)_L$ are $B_{\mu\nu}$ and $W_{\mu\nu}^a$, respectively, 
with $g_1$ and $g_2$ the corresponding gauge couplings. 
The dimensionless Wilson coefficients $\Gamma_B^{u,d}$ and $\Gamma_W^{u,d}$ quantify the coupling strengths of the light-quark dipole interactions. 
After expanding Eq.~\eqref{eq:Leff}, we define for convenience the light-quark dipole couplings to the photon and $Z$ boson as 
$\Gamma_\gamma^{d,u} = \pm \Gamma_W^{d,u}-\Gamma_B^{d,u}$ and 
$\Gamma_Z^{d,u} = \pm c_W^2\Gamma_W^{d,u} + s_W^2\Gamma_B^{d,u}$, 
where the ``$\pm$'' refers to the $d$ or $u$ quark, respectively,
and $(s_W, c_W) \equiv (\sin\theta_W, \cos\theta_W)$, with $\theta_W$ being the weak mixing angle. 

Since the dipole interactions flip the quark helicities, they contribute to $W_{-+}^q$ at $\order{1/\Lambda^2}$ through interference with the SM. 
This quark-helicity-flipping effect is unique to dipole operators in Eq.~\eqref{eq:Leff} and is not present in any SM or other dimension-six SMEFT operators.
Hence, the transverse spin effect of the final-state quark is a clean signature of the dipole operators.
Without observing this effect, the off-diagonal information $W_{-+}^q$ would be missing, 
leaving merely the measurement of the unpolarized rate 
$C_q = \tr(W^q) = W_{++}^q + W_{--}^q$. 
This receives no contribution from the interference of SM and dipole operators, so that the dipole interaction 
would only contribute quadratically at $\order{1/\Lambda^4}$.

It is easy to see from Eq.~\eqref{eq:Leff} that
at the leading power of $1/\Lambda^2$, the helicity amplitudes depend on the dipole operators in a rather simple way,
$ (\M_{+-}, \M_{-+}) \propto (\Gamma_{\gamma, Z}^q, \Gamma_{\gamma, Z}^{q*})$, 
and so does the off-diagonal element of the quark density matrix, 
\beq
	W^q_{-+} \propto \M_{+-} \M^*_{++} + \M_{--} \M^*_{-+} \propto \Gamma^q_{\gamma, Z},
\eeq
where we have omitted the dummy indices of the electron helicities.
Hence, at $\mathcal{O}(1/\Lambda^2)$, one can write
\beq
	W^q_{-+} = w^q_{\gamma} \, \Gamma^q_{\gamma} + w^q_Z \, \Gamma^q_Z.
\eeq
In contrast, the unpolarized rate $C_q = W^q_{++} + W^q_{--}$ does not depend on $\Gamma_{\gamma, Z}^q$ at $\mathcal{O}(1/\Lambda^2)$. At the LO, the coefficients $w_{\gamma,Z}^q$ are real. Comparing to Eq.~\eqref{eq:hard}, we obtain 
\begin{align}
	s^x_q &= \frac{2}{C_q} \left( w^q_{\gamma} \, \Re\Gamma^q_{\gamma} + w^q_Z \, \Re\Gamma^q_Z \right), \nn\\
	s^y_q &= \frac{2}{C_q} \left( w^q_{\gamma} \, \Im\Gamma^q_{\gamma} + w^q_Z \, \Im\Gamma^q_Z \right).
\label{eq:w-G}
\end{align}

The imaginary parts of $\Gamma_{\gamma}^q$ and $\Gamma_Z^q$ break the parity and $CP$ symmetries for both the $\gamma$ and $Z$ interactions.
It is easy to show~\cite{Yu:2023shd} that for the $2\to2$ electron-quark scattering, $s^x_q$ is a parity-odd quantity while $s^y_q$ is parity-even.
Therefore, both the $w^q_{\gamma}$ and $w^q_Z$ need to involve parity-odd structures. 
Till now we have been considering unpolarized scattering, for which the only parity-odd structure is the parity-violating $Z$ interaction $g^{e, q}_A$ in the SM.
The EIC will also allow a longitudinal polarization $\lambda_{\ell}$ of the electron, which amounts to modifying the definition of $W^q$ in Eq.~\eqref{eq:hard}
by inserting a factor $(1 + \lambda_{\ell} \lambda_e)$ in the sum of $\lambda_e$.
Both cases are effectively inserting a $\gamma_5$ in the fermion trace when evaluating Eq.~\eqref{eq:hard}.
The other possibility is to have a longitudinally polarized proton beam. 
That, however, is suppressed by the helicity parton density so is not considered in this work.

The contributions to the $W^q_{\lambda\lambda'}$ can be divided into three channels:
the amplitude square of a photon ($\gamma\gamma$) or $Z$ exchange ($ZZ$),
and the interference of a photon and a $Z$ exchanges ($\gamma Z$).
At the EIC energy, the $\gamma\gamma$ channel dominates, followed by the $\gamma Z$ channel,
while the $ZZ$ channel can be neglected to a good approximation~\cite{Yan:2021htf}. 
In Eq.~\eqref{eq:w-G}, the $w_{\gamma}^q$ contains contributions from both the $\gamma\gamma$ and $\gamma Z$ channels,
where the dominant contribution in the $\gamma\gamma$ channel requires a nonzero $\lambda_{\ell}$.
On the other hand, the $w^q_Z$ contains both the $\gamma Z$ and $ZZ$ channels, 
where the major contribution in the $\gamma Z$ channel is proportional to the parity-odd electron-$Z$ coupling $g^e_A$~\cite{Yan:2021htf}
and does not require a $\lambda_{\ell}$. 
Therefore, the electron polarization provides an efficient way to distinguish the two couplings $\Gamma^q_{\gamma}$ and $\Gamma^q_Z$.
With a sizable $\lambda_{\ell}$, the $\bm{s}_{T, q}$ is more sensitive to $\Gamma^q_{\gamma}$, 
whereas without $\lambda_{\ell}$, it becomes more sensitive to $\Gamma^q_Z$.

\section{Numerical results and discussion}
Below, we study the potential of the  EIC for probing the dipole operators by measuring azimuthal asymmetries in the inclusive $\pi^+\pi^-$ dihadron productions.
For simplicity, we integrate over $(x, y, z, M_h)$ in Eq.~\eqref{eq:diXsec} in the region 
$x \in [0.03,0.9]$, $y \in [0.3,0.95]$, $z\in [0.19,0.99]$, and $M_h\in [0.28,2.05]~{\rm GeV}$,
leaving a more detailed binning to future experimental analyses.
Our choice of the kinematic ranges in terms of variables $x$ and $y$ ensures $Q > 10$ GeV to satisfy the factorization condition in Eq.~\eqref{eq:diXsec}, which also aligns well with the accessible kinematic range at the EIC~\cite{Accardi:2012qut}.
The ranges of $z$ and $M_h$ coincide with those in the recent JAM global analysis of the DiFFs~\cite{Pitonyak:2023gjx,Cocuzza:2023oam,Cocuzza:2023vqs},
whose results we take for the $D_{h_1 h_2 / q}$ and $H_{h_1 h_2 / q}$ in our numerical computation. 
These DiFFs satisfy the following flavor relations as consequences of isospin and charge-conjugation symmetries:
\begin{align}
	&D_{\pi^+\pi^-/u} = D_{\pi^+\pi^-/d}, \;\;
	&H_{\pi^+\pi^-/u} = -H_{\pi^+\pi^-/d},	\;\;\nn\\
	&D_{\pi^+\pi^-/q} = D_{\pi^+\pi^-/\bar{q}}, \;\;
	&H_{\pi^+\pi^-/q} =- H_{\pi^+\pi^-/\bar{q}},\;\;\nn\\
 	&H_{\pi^+\pi^-/(s, \bar{s}, c, \bar{c}, b, \bar{b})} = 0.\;\;
\label{eq:diff-isospin}
\end{align}
As a result, only the $u$ and $d$ quarks can contribute to the azimuthal modulations in Eq.~\eqref{eq:diXsec}.
After integration, the cross section becomes single differential in $\phi_R$,
\begin{align}
	\frac{2\pi}{\sigma_{\rm tot}} \frac{d\sigma}{  d\phi_R}
	= 1 + A_R \sin\phi_R + A_I \cos\phi_R + \order{1/\Lambda^4},
\label{eq:dis}
\end{align}
where $A_R$ and $A_I$ are obtained by averaging the transverse spins $s^x_q$ and $s^y_q$ in Eq.~\eqref{eq:diXsec}, respectively,
together with the spin analyzing power $H_{h_1 h_2 / q} / D_{h_1 h_2 / q}$.
These coefficients can be extracted simultaneously from the azimuthal asymmetry observables,
\begin{align}
	A_{UD}
	&= \frac{\sigma(\sin\phi_R>0) - \sigma(\sin\phi_R<0)}{\sigma(\sin\phi_R>0) + \sigma(\sin\phi_R<0)}
	= \frac{2}{\pi}A_R,\nn\\
	A_{LR}
	&= \frac{\sigma(\cos\phi_R>0) - \sigma(\cos\phi_R<0)}{\sigma(\cos\phi_R>0) + \sigma(\cos\phi_R<0)}
	= \frac{2}{\pi} A_I,
\label{eq:SSA}
\end{align}
where $\sigma(\sin\phi_R \lessgtr 0)$ represents the integrated cross section with $\sin\phi_R \lessgtr 0$, etc.
They can be referred to as ``up-down'' and ``left-right" asymmetries with respect to the lepton plane and $\hat{y}$ axis in Fig.~\ref{Fig:Geo}, respectively.

As evident from Eq.~\eqref{eq:w-G}, the measurements of $A_{UD}$ and $A_{LR}$ can constrain the real and imaginary parts of the dipole couplings, respectively.
To assess the constraining power at the EIC, we calculate the statistical uncertainties of these asymmetries using the formula
\begin{equation}
	\delta A_{LR, UD}
	=\sqrt{\frac{1 - (A_{LR, UD})^2}{N_{\rm events}}}
	\simeq \frac{1}{\sqrt{N_{\rm events}}},
\label{eq:stats}
\end{equation}
where $N_{\rm events}$ represents the number of selected total events after applying kinematic cuts for a given collider energy and integrated luminosity. 
In the second step of Eq.~\eqref{eq:stats}, we have used $A_{LR, UD} \simeq 0$ in the SM. 
The systematic uncertainties in the asymmetries are canceled in the ratios and can be disregarded in this study.
Within the adopted kinematic region, we further estimate that the relative uncertainties in the spin asymmetries arising from the DiFF are below 20\%, with statistical errors being the dominant contribution, so that the DiFF-induced uncertainties have negligible impact on our conclusions and are expected to decrease with future data from the EIC and lepton colliders. Since maximizing the integrated luminosity $\mathcal{L}$ has a more significant impact on the sensitivity for probing the SMEFT effects,  
compared to a slight increase in collider energy $\sqrt{s}$ at the EIC~\cite{Boughezal:2022pmb,Boughezal:2023ooo,Wang:2024zns}, 
we take  $\sqrt{s}=105~{\rm GeV}$ 
as it is anticipated that the highest integrated luminosity will be achieved~\cite{Accardi:2012qut,AbdulKhalek:2021gbh},
using $\mathcal{L}=1000~{\rm fb}^{-1}$ as a benchmark.

\begin{figure}
\centering
	\includegraphics[scale=0.33]{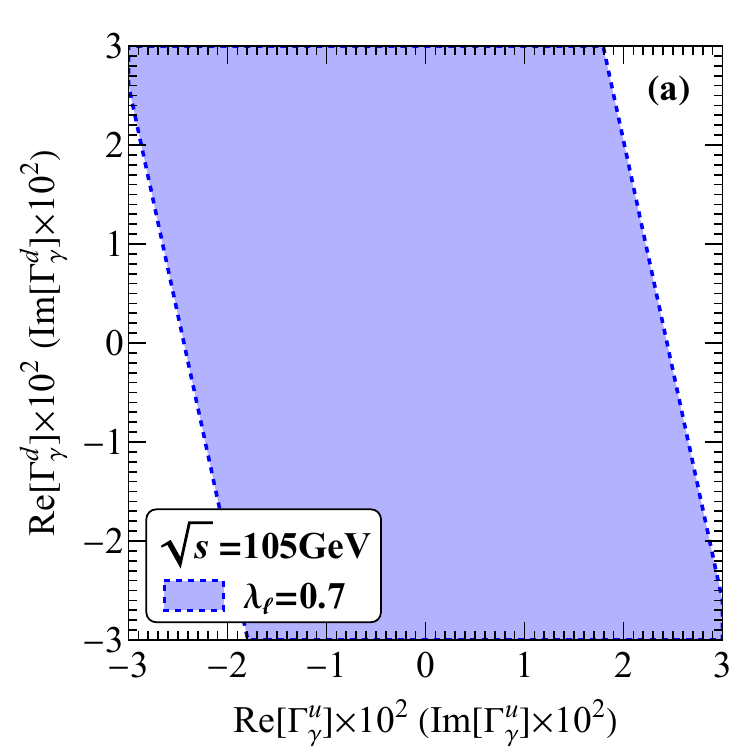}
	\includegraphics[scale=0.33]{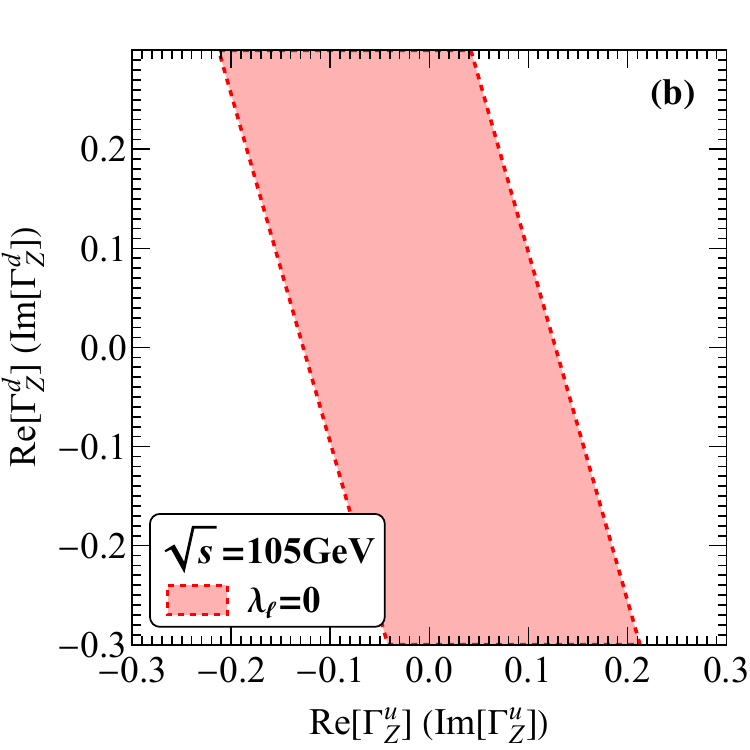}
	\includegraphics[scale=0.33]{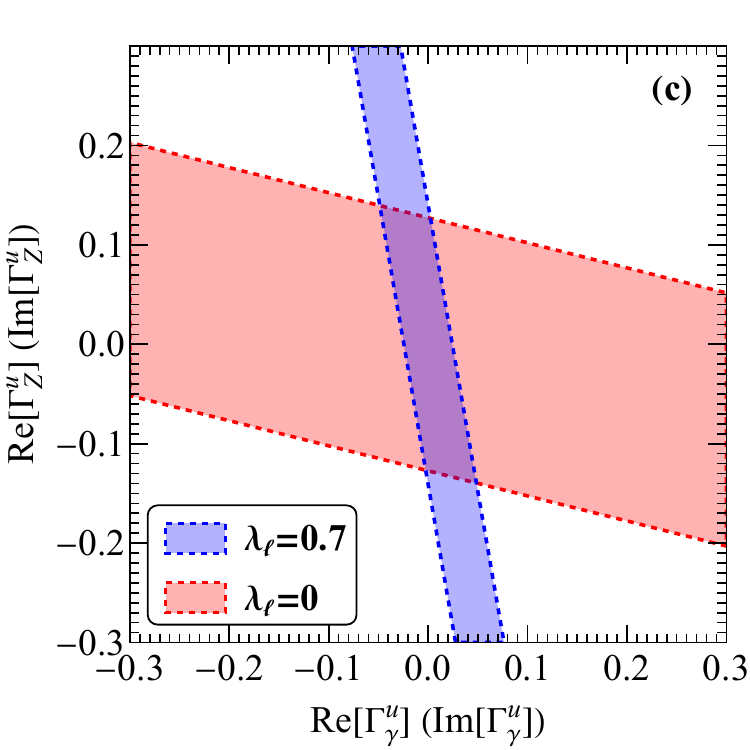}
	\includegraphics[scale=0.33]{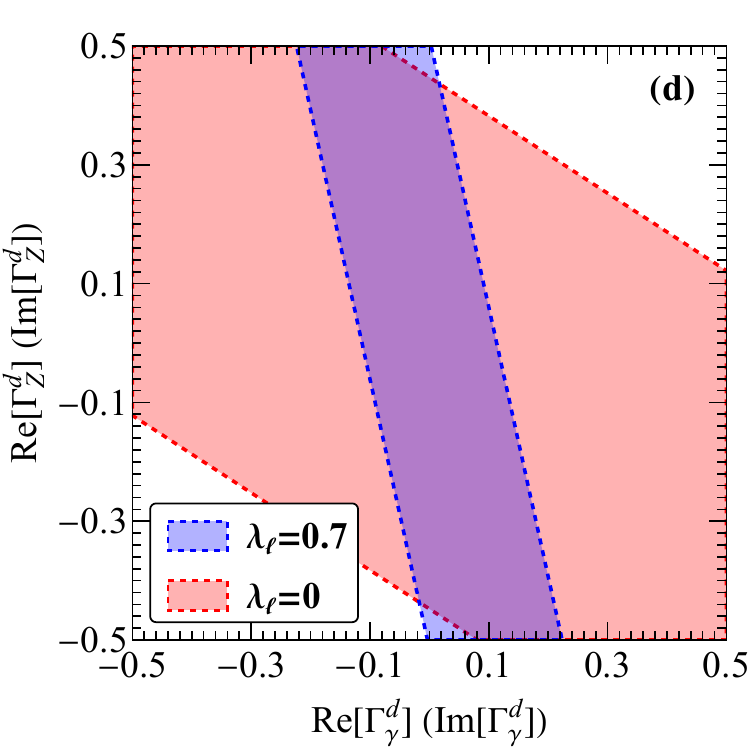}
	\caption{Expected constraints on the real and imaginary parts of the light-quark dipole couplings 
		(a) $\Gamma_{\gamma}^{u,d}$ with $\lambda_\ell=0.7$, 
		(b) $\Gamma_{Z}^{u,d}$ with $\lambda_\ell=0$,
		(c) $\Gamma_{\gamma,Z}^{u}$ assuming $\Gamma_{\gamma,Z}^{d}=0$, and 
		(d) $\Gamma_{\gamma,Z}^{d}$ assuming $\Gamma_{\gamma,Z}^{u}=0$, 
  from azimuthal asymmetries $A_{UD}$ and $A_{LR}$.}
\label{Fig:Limit}
\end{figure}

The projected constraints on the dipole couplings $\Gamma_{\gamma,Z}^{u,d}$ are presented in Fig.~\ref{Fig:Limit} at the 68\% confidence level.
The linear boundaries are due to the linear dependence of the asymmetries on these couplings, probing linear combinations of $\Gamma_{i}^{u}$ and $\Gamma_{i}^{d}$ ($i=\gamma,Z$) as well as that of $\Gamma_{\gamma}^{j}$ and $\Gamma_{Z}^{j}$ ($j=u,d$). 
The primary constraint is on the photon dipole couplings $\Gamma_{\gamma}^{u,d}$ when we take $\lambda_{\ell} = 0.7$ for a polarized electron beam, 
whereas it shifts to the $Z$ dipole couplings $\Gamma_{Z}^{u,d}$ in the case of $\lambda_\ell = 0$ for an unpolarized electron beam. 
As expected from Eq.~\eqref{eq:SSA}, the real and imaginary parts are constrained simultaneously and with equal power,
with typical bounds of $\mathcal{O}(10^{-2})$ for $\Gamma_{\gamma}^{u,d}$ and $\mathcal{O}(10^{-1})$ for $\Gamma_{Z}^{u,d}$, 
when considering one of $( \Gamma_{\gamma}^{u}, \Gamma_{\gamma}^{d}, \Gamma_{Z}^{u}, \Gamma_{Z}^{d} )$ at a time.
Due to the larger electric charge and parton density of the up quark compared to the down quark, 
the sensitivity to $\Gamma_{\gamma,Z}^{u}$ is stronger than $\Gamma_{\gamma,Z}^{d}$.
We have checked that narrowing the $z$ bin in the integration of Eq.~\eqref{eq:diXsec} 
to the range where the DiFFs are most precisely constrained does not alter our qualitative result.
Nevertheless, we expect the above constraints to be further improved quantitatively with optimized kinematic bins and cuts and colliding energy and luminosity.

With only one dihadron channel $\pi^+ \pi^-$, one cannot constrain all of these couplings at the same time, 
so our results ought to be combined with other methods and improved with more dihadron production channels in the future~\cite{Wen:2024nff}.
In practice, the photon dipole coupling $\Gamma_{\gamma}^{u, d}$ can be almost exclusively constrained with the setting $\lambda_{\ell} = 0.7$;
after that, the $Z$ dipole coupling $\Gamma_Z^{u, d}$ can be well constrained with $\lambda_{\ell} = 0$.
Hence, for simplicity, we turned off $\Gamma_Z^{u, d}$ in Fig.~\ref{Fig:Limit}(a) and $\Gamma_{\gamma}^{u, d}$ in Fig.~\ref{Fig:Limit}(b), while switching off $\Gamma_{\gamma,Z}^{d}$ in Fig.~\ref{Fig:Limit}(c) and $\Gamma_{\gamma,Z}^{u}$ in Fig.~\ref{Fig:Limit}(d) for combining these two configurations of electron polarization. 

In the recent SMEFT analyses, constraints on the light quark dipole couplings typically come from the Drell-Yan process at the LHC,
which constrains $\Gamma_{\gamma, Z}^q$ at $\mathcal{O}(1/\Lambda^4)$,
with sensitivity weaker than our results by several times to an order of magnitude, even when considering only one operator at a time~\cite{Boughezal:2021tih}.
Light-quark dipole operators also contribute to the electric and magnetic dipole moments of nucleons and can therefore be constrained by low-energy precision measurements, leading to bounds of $|{\rm Im}\,\Gamma_\gamma^q| \sim \mathcal{O}(10^{-6})$ and $|{\rm Re}\,\Gamma_\gamma^q| \sim \mathcal{O}(0.1)$~\cite{Liu:2017olr,ParticleDataGroup:2022pth}. By contrast, the dipole coupling to the $Z$ boson can only be probed indirectly through renormalization effects, resulting in significantly weaker constraints. It should be noted, however, that extracting limits on the underlying quark-level dipole couplings from low-energy observables is inherently theory dependent and relies on nonperturbative hadronic inputs, which introduce sizable uncertainties. Additional constraints from electroweak precision observables, and
differential cross section measurements of Higgs boson production via vector boson fusion, diboson production, 
and associated production of weak and Higgs bosons,
are comparable or even worse than those from the Drell-Yan process~\cite{Escribano:1993xr, Alioli:2018ljm, daSilvaAlmeida:2019cbr}. 
None of these methods thus far could separate the dipole operators from other SMEFT operators. 
Furthermore, they lack the sensitivity to distinguishing between the real and imaginary parts of the dipole couplings. 
This highlights the unique opportunity provided by the dihadron observable at the EIC. 

With transverse spins of the electron and proton, more observables could be constructed for probing the dipole operators at tree level.
We leave this study to future works, noting only that the $\gamma\gamma$ channel typically requires a  nonvanishing beam or target longitudinal polarization to make a contribution,
which is likely to improve the sensitivity of the EIC reported in Ref.~\cite{Boughezal:2023ooo}.

\section{ Conclusions}
In summary, light-quark dipole operators are difficult to be constrained due to both the confinement of quarks and their chiral-odd nature. In this work, we propose a novel approach to sensitively probe them at the leading power of $1/\Lambda^2$ 
by observing azimuthal asymmetries in the inclusive $\pi^+\pi^-$ dihadron productions in the deep inelastic electron-proton scattering at the upcoming EIC.
The transversely polarized quark produced from the interference of the dipole operators with the SM 
is connected to the dihadron's azimuthal asymmetry through a chiral-odd dihadron fragmentation function, obtained from a global fit. This approach does not require a polarized proton 
and is exclusively sensitive to the light-quark dipole couplings,
and free of contamination from other new physics operators.
It enables a simultaneous determination of both the real and imaginary parts of dipole couplings, 
offering a new way to probe $CP$-violating effects at high energies.
We have demonstrated that this new observable can improve the constraints on these dipole couplings by an order of magnitude as compared to usual methods discussed in the literature.
In this way, our framework opens new avenues for probing new physics at the precision frontier, extends the electroweak discovery potential of the EIC and future lepton colliders~\cite{Gonderinger:2010yn,Boughezal:2020uwq,Cirigliano:2021img,Anderle:2021wcy,AbdulKhalek:2021gbh,Yan:2021htf,Li:2021uww,Davoudiasl:2021mjy,Liu:2021lan,AbdulKhalek:2022hcn,Yan:2022npz,Boughezal:2022pmb,Zhang:2022zuz,Batell:2022ogj,Boughezal:2023ooo,Davoudiasl:2023pkq,Balkin:2023gya,Wang:2024zns,DaRold:2024ram,Ellis:2025ghl,Ma:2025fmy,Cao:2025qua,Cao:2025wfg,He:2026pxa}, and establishes a direct connection between hadron structure studies and electroweak precision physics through QCD spin phenomena.

\vspace{3mm}
We thank C.~Cocuzza, A.~Prokudin, and N.~Sato for sharing the dihadron fragmentation function code and helpful discussions.
Xin-Kai Wen is supported in part by the National Science Foundation of China under Grants Nos.~12235001,~12342502 and~12547174. 
Bin Yan is supported in part by the National Science Foundation of China under Grant No.~12422506, the IHEP under Contract No.~E25153U1 and CAS under Grant No.~E429A6M1.
Zhite Yu is supported in part by the U.S. Department of Energy (DOE) Contract No.~DE-AC05-06OR23177, 
under which Jefferson Science Associates, LLC operates Jefferson Lab 
and C.-P.~Yuan is supported by the U.S. National Science Foundation under Grant No.~PHY-2310291.

\noindent {\bf Conflict of Interest:} The authors declare that they have no conflict of interest.

\bibliographystyle{apsrev}
\bibliography{reference}

\end{document}